# Parallelizing the MARS15 Code with MPI for Shielding Applications*†

M.A. Kostin  and  N.V. Mokhov

*Fermilab, P.O. Box 500, MS 220, Batavia, Illinois 60510-0500, USA*

April 27, 2004

### Abstract

The MARS15 Monte Carlo code capabilities to deal with time-consuming deep penetration shielding problems and other computationally tough tasks in accelerator, detector and shielding applications, have been enhanced by a parallel processing option. It has been developed, implemented and tested on the Fermilab Accelerator Division Linux cluster and network of Sun workstations. The code uses MPI. It is scalable and demonstrates good performance. The general architecture of the code, specific uses of message passing, and effects of a scheduling on the performance and fault tolerance are described.

\* Presented paper at the *10th International Conference on Radiation Shielding*, Funchal (Madeira), Portugal, May 9-14, 2004.
† This work was supported by the Universities Research Association, Inc., under contract DE-AC02-76CH03000 with the U.S. Department of Energy.

# PARALLELIZING THE MARS15 CODE WITH MPI FOR SHIELDING APPLICATIONS

M.A. Kostin* and N.V. Mokhov

*Fermi National Accelerator Laboratory, P.O. Box 500, MS 220, Batavia, Illinois 60510-0500, USA*
*Corresponding author: phone +1-630-840-8460, fax +1-630-840-6039, e-mail: kostin@fnal.gov

**Abstract** – The MARS15 Monte Carlo code capabilities to deal with time-consuming deep penetration shielding problems and other computationally tough tasks in accelerator, detector and shielding applications, have been enhanced by a parallel processing option. It has been developed, implemented and tested on the Fermilab Accelerator Division Linux cluster and network of Sun workstations. The code uses MPI. It is scalable and demonstrates good performance. The general architecture of the code, specific uses of message passing, and effects of a scheduling on the performance and fault tolerance are described.

INTRODUCTION

Shielding applications are known to be time consuming. A usual way to overcome a low rate of statistics accumulation in such applications is to run many jobs with different initial seeds for random number generators and average their results. The procedure, however, may result in biased estimates for values of interest and their errors if the results of the jobs are not statistically significant.

There is a natural way to deal with the problem. A code can be created that combines the local jobs in one system, collects intermediate statistics from the jobs, and evaluates the final results. The job integration is done via a so-called *middleware*. This approach was recently implemented in the MARS15 code[1,2] by means of the Message Passing Interface (MPI) libraries[3].

CHOICE OF MIDDLEWARE

Several candidates were considered for the middleware: MPI[3], CORBA[4], sockets and PVM[5]. CORBA provides extensive functionality and is especially appropriate for distributed Object-Oriented applications. The drawbacks of CORBA is that it is relatively hard to use, and its communication overhead may be significant. Sockets, on the other hand, involve little overhead for communications but much of the necessary high-level functionality is absent. PVM has a long and successful history. The MCNP collaboration[6], however, has recommended MPI over PVM provided substantial experience with both the packages. MPI has the following advantages. It is a standard for programming parallel systems, available on machines of all architectures, i.e. massive parallel systems, clusters and networks of workstations. The performance of MPI is optimized by many vendors for their systems. At least one of the MPI implementations (namely LAM[7]) is Grid capable, which may be important for future applications of the MARS code. Also, LAM uses the TCP/IP protocol that imposes virtually no communication overhead. Another advantage of MPI is that its functionality seems to match data structures and structure of the MARS code quite well. Summarizing, MPI seems to be the most appropriate choice among all the other considered middlewares.

CODE ARCHITECTURE

The general architecture of the parallel MARS15 code consists of one *master* process and an arbitrary number of *slave* processes. All the processes replicate the entire geometry of a studied system. The parallelization is job-based, i.e. processes are running independently with different initial seeds for the random number generator. The master process collects intermediate results from the slaves from time to time according to a *scheduling* algorithm, and computes the final results when a required total number of events has been processed. Besides performing the control task, the master also runs event histories. This is especially important for systems with a small number of processors. Since the processes are loosely coupled, one expects good scalability, and *load balancing* is not an issue.

The slave processes are passive, they do not perform self-scheduling tasks. Information exchange is initiated by the master whenever the scheduling algorithm decides. All the slaves are inquired consecutively according to their *rank*. In order to avoid possible interference with running event histories, the slaves probe the signals from the master in *asynchronous* mode. A slave starts processing the next event if no signal from the master is received at the time of the probing. If such a signal exists, then the slave sends back a number of locally processed events. The master makes an estimate of the total number of events processed by all the processes including that new number. The slave process is terminated if the estimate is in excess of the requested total number of events.

Otherwise, all the intermediate information is transferred to the master and the slave process continues. All communications are performed in MPI *standard* mode except for the signal probing mentioned above. The order of all the corresponding 'send' and 'receive' functions is carefully matched in order to avoid a *deadlock*.

Information exchange between two processes has two stages. At the first stage, all the relevant MARS arrays are sent. There are three general methods of sending the data in the MPI:
1. Sending array elements positioned contiguously in memory.
2. Sending with the packing and unpacking subroutines.
3. Sending with *derived types*.

The second and third methods work with data located in arbitrary places. MARS arrays are positioned in a number of common blocks, therefore the arrays are not in contiguous memory. The first method can only be used for one array at a time. This would lead to excessive communications since each array should be sent separately. The second method involves some overhead for packing the arrays into a buffer every time a slave sends the information. The third method is an optimal one for our case. An overhead for type creating occurs only once at the beginning of the run. Later on, all the arrays pertained to the new type are sent by a single send call.

The last two methods can be mixed in a send/receive session as long as *type signatures* of sent and received messages are same. This is used for receiving the data by the master. If the same derived type were used by the master to receive the information then all the arrays in the master's memory would be overwritten. A buffer is required to store the data received from the slave processes. The FORTRAN'77 standards do not provide utilities for dynamic memory allocation. Even though many vendors provide those for their implementations, the free distributed and commonly used *gnu* FORTRAN'77 compiler, g77, does not do that. The maximal possible number of zones in a studied system has to be hardwired, and all the arrays that keep the data associated with the zones must have fixed sizes. Therefore, one can safely fix the size of the exchange buffer to the maximal possible integrated size of all those arrays, that is ~40 MB.

At the second stage, the slave processes send accumulated HBOOK histograms. The code reuses the same buffer mentioned above to do this. The histograms are sent one by one in the current version of the code. It might be possible to further optimize the performance by packing as many histograms as possible into the exchange buffer and reducing the number of send calls. A histogram can be created at any time, even between exchange sessions. The master checks dynamically whether a received histogram with a given ID already exists. It will be created if it is not found among the master histograms.

SCHEDULING

Scheduling is an important issue which is directly related to performance, scaling and fault tolerance.

Communications are quite expensive for the current generation of *commodity clusters*. Therefore, a common sense approach is important to design an effective code: the less communications the better. In the most extreme case, the result exchange (*rendezvous*) would happen only once at the end of computation. Moreover, performance can be good only if communication time, $T_m$ (the subscript 'm' stands for 'messages'), is much smaller than computation time, $T_c$. On the other hand, the rendezvous must be frequent enough in order to provide some fault tolerance. It is important to be able to restart computation from the last checkpoint if a system failure occurs. Even though this functionality has not been implemented in MARS yet, the scheduler compromises between these two time requirements in order to satisfy them.

The scheduler in the MARS15 code decides when to suspend the computation and start a rendezvous. The decision is based on the knowledge of an estimated time needed for a rendezvous and to process one event, $T_1$. The master process may wait for a response from a slave for a long time during rendezvous in a case of long histories. The waiting time may significantly prolong the rendezvous. For the code to be effective, a time between rendezvous has to be significantly longer than $max\{T_m, T_1\}$.

As described above, a slave process is terminated at a rendezvous if the number of locally processed histories combined with the number of events already collected by the master is in an excess of the total number of requested events. This would most likely to happen when the jobs are close to their end. Time to a next rendezvous has to be shortened to avoid that and use the resources more effectively. In the opposite case, the master will have to process the rest of histories by itself. This may lead to a sizeable computation time increase if a number of terminated jobs is large and the balance of events is still significant. Two previous time conditions must be corrected for this effect. The time till a next rendezvous is calculated according to the formulae

$$T = \min\{100 \times \max\{T_1, T_m\}, 0.8 \times T_{end}, 1\ hr\},$$

where $T_{end}$ is an estimated time to the end of all calculations. The requirement of 1 hour is based on a human factor. The exchange time must not exceed a sizeable fraction of a working day. This is to let people deal with potential problems in their codes.

Fig. 1 illustrates a communication time scheme. The first rendezvous is occurring in a fixed time period of 10 sec. The fixed time period is needed to calculate trial values for $T_1$, $T_m$ and $T_{end}$. These values are calculated at the end of each rendezvous later on.

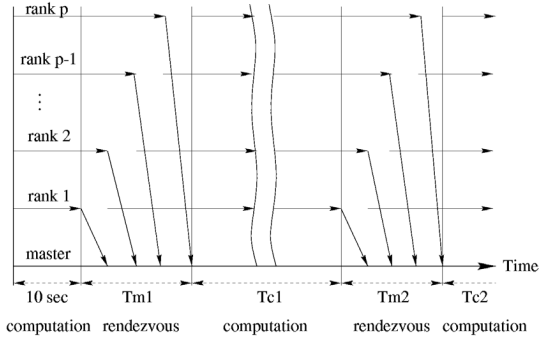

Figure 1. Communication time scheme.

PERFORMANCE TEST

A performance test has been conducted on the Fermilab Accelerator Division Linux cluster. The cluster consists of 32 dual CPU nodes. Each CPU is a 1.4 GHz AMD Athlon 1600+ processor with 256 kbytes of cache. The nodes are equipped with 1 Gbytes of memory and connected via a 100 Mbits/sec network. LAM MPI is installed on the cluster. Job management is handled by the Portable Batch System (PBS)[7].

The test was conducted on a MARS model of the secondary beam line of the Fermilab fixed target experiment E907 (another name is MIPP - Main Injector Particle Production Experiment)[9]. The model includes last 130 m of the beam line and beam enclosure with concrete shielding and soil around it. An elevation and horizontal views of the system are shown in Figs.2 and 3. The model contains ~2,000 geometry zones. An event in the model starts with transport of a 120 GeV/c proton. Then the proton interacts with a primary copper target, and secondary particles are transported further and their interactions in material of the beam line elements and shielding are simulated.

The performance test has measured the *speedup*, $S_N$, and *efficiency* of the code, $E_N$, as functions of the used number of processors, N. The speedup is defined as a ratio of the time spent by one processor to perform a given job, $t_1$, over the time $t_N$ required for N processors to execute the same job, $S_N=t_1/t_N$. The efficiency of the code is defined as $E_N=t_1/(N \times t_N)$. For this test, $t_1$ and $t_N$ were averaged times to process one event. The results of the performance test are presented in Figs. 4 and 5. A number of requested events was 10,000 per processor per each point on the plot. The jobs took ~90 min to complete. Only one processor per node was used during the test. The test has demonstrated an almost linear speedup. The test could not be conducted on a large number of processors due to the limited resources.

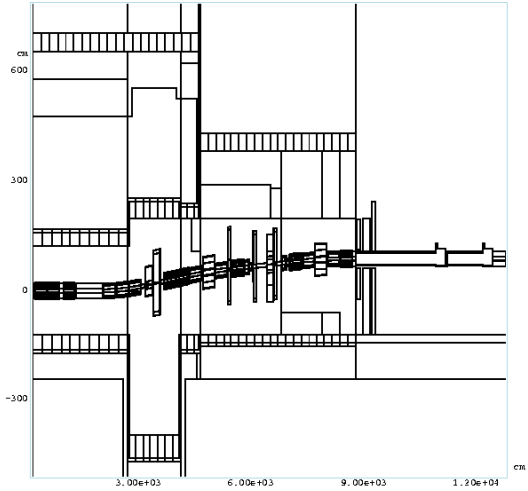

Figure 2. Elevation view of the MIPP secondary beam line.

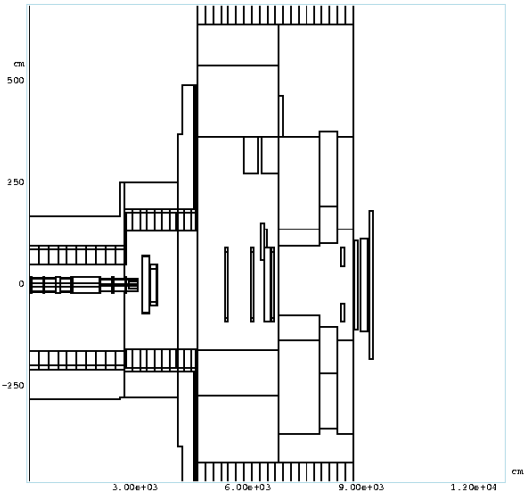

Figure 3. Horizontal cross-section of the MIPP beam line enclosure.

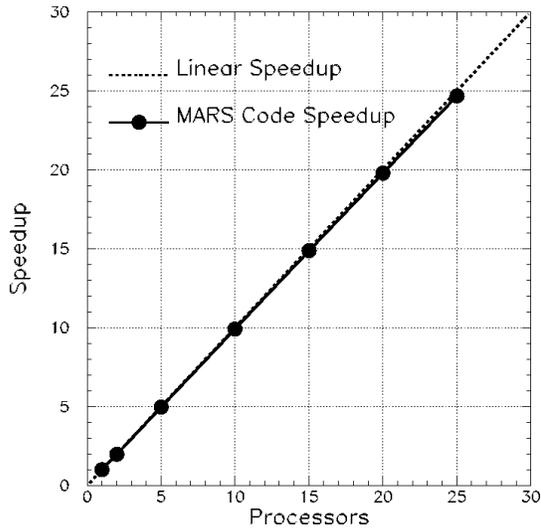

Figure 4. MARS15 code speedup, $S_N$.

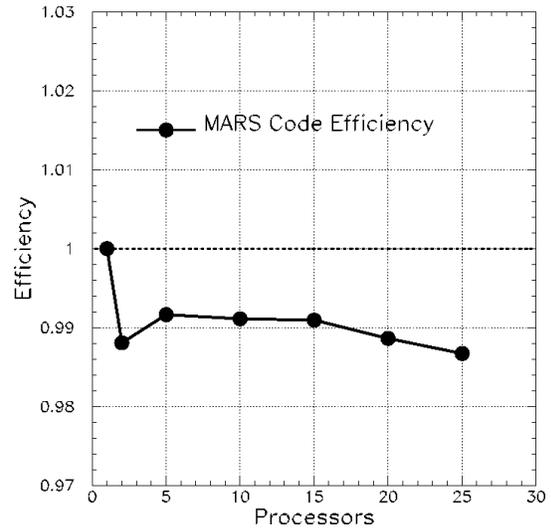

Figure 5. MARS15 code efficiency, $E_N$.

## CONCLUSIONS

Code parallelization has been implemented in MARS15 using MPI. The code is expected to be scalable because the processes are weakly coupled. There is no issue of the load balancing due to the same reason. The scheduler is an important part of the code. It comprises time conditions to provide efficiency and fault tolerance. A performance test conducted on the Fermilab Accelerator Division Linux cluster has shown a good speedup. There is potential for further improvement of the code.

Only computation and histograming are supported in the current version of the parallel MARS15 code. All other options such as graphics, event dumping and so on shall be used with a sequential run on one processor. All the nodes in a cluster must have some output capabilities because the code writes multiple files. Therefore, the code most likely will not run on disc-less clusters. However, only the master node needs to have the output capabilities to print out the results. The current version of the code also assumes an unlimited input, i.e. each node must be able to read in the configuration files.

## ACKNOWLEDGEMENTS

We are thankful to J.-F. Ostiguy for fruitful discussions and constructive comments.